\documentclass[apj]{emulateapj}
\usepackage{graphicx,amssymb,amsmath,epsfig}
\usepackage{natbib}
\usepackage{subfigure}

\def\epsr{\epsilon_{rad} }

\begin{document}
\normalsize

\author{Gilad Svirski and Ehud Nakar}
\affil{Raymond and Beverly Sackler School of Physics \&
Astronomy, Tel Aviv University, Tel Aviv 69978, Israel}

\title{Spectrum and light curve of a supernova shock breakout through a thick Wolf-Rayet wind}

\begin{abstract}
Wolf-Rayet  stars are  known  to eject winds.  Thus, when  a
Wolf-Rayet  star explodes  as a  supernova, a  fast, $>30,000$  km/s, shock  is
expected to be driven  through a wind. We study the  signal expected from a
fast  supernova shock  propagating through  an optically thick  wind, and  find that  the
electrons behind  the shock driven  into the  wind are cooled  efficiently, by
inverse  Compton  over soft  photons  that  were  deposited by  the  radiation
mediated shock that crossed the  star. Therefore, the bolometric luminosity is
comparable to the  kinetic energy flux through the shock,  and the spectrum is
found to be a power-law, which slope  and frequency range depend on the number
flux of soft photons available for cooling. Wolf-Rayet supernovae that explode
through  a thick  wind have  a high  flux of  soft photons,  producing a  flat
spectrum, $\nu F_{\nu}=Const$,  in the X-ray range  $0.1\lesssim T\lesssim 50$
keV. As the shock expands into an optically thin wind, the soft photons are no
longer able to  cool the shock that  plows through the wind,  and the bulk of the emission
takes  the  form of  a  standard  core-collapse  supernova (without  a  wind).
However, a  small fraction of the  soft photons is upscattered  by the shocked
wind     and    produces     a    transient     unique    X-ray     signature.
\end{abstract}

\section{Introduction}

The scenario of a supernova (SN) exploding through a thick wind has been the focus of many recent papers \citep[e.g.][]{Ofek+2010,Balberg_Loab2011,Chevalier_Irwin2011,Murase+2011,Katz+2011,Moriya_Tominaga2011,Chevalier_Irwin2012,Svirski+2012,Balberg+2012}.
The events discussed in these papers are mostly classified as type IIn SNe, and although they only account for a few percents of the observed SNe (e.g. \citealt{Li+2011}), they provide a unique probe for the progenitor's pre-explosion composition and mass-loss history. In some cases, such information may even reveal an unexpected new SN progenitor type (e.g., a luminous blue variable, \citealt{Kiewe+2012,Mauerhan+2013}) or a possible causal connection between a pre-explosion mass-loss burst and the SN explosion \citep{Ofek+2013}.


The study  of SNe  that interact  with a thick  wind has  focused thus  far on
interaction  timescales of  days or  longer, currently  the most  probable for
detection. However,  some Wolf-Rayet (WR)  SNe are  naturally expected  to interact
with a thick wind, and this interaction timescale should be much shorter than
a   day.    Here   we   study    this   common   yet    unexplored   scenario.

If, due to a massive mass loss, a SN progenitor is surrounded by a wind of an optical depth $>c/v$, where $c$ is the speed of light and $v$ is the SN shock velocity, then the shock does not break out near the stellar edge: After crossing the star's envelope, the radiation dominated shock continues into the wind, while a reverse shock is driven into the SN ejecta.
The interaction of the expanding ejecta with the thick wind leads to a very luminous SN, that breaks out once the optical depth of the wind ahead of the shock decreases to $c/v$.
Following the breakout, the radiation mediated shock is replaced by a narrower collisionless shock and a layer of hot shocked electrons, initially with a temperature $T_h\gtrsim 60$ keV (hereafter, a temperature $T$ denotes an energy $k_BT$), forms behind the shock \citep{Katz+2011,Murase+2011}. The interplay of this layer with the surrounding colder medium, mediated by the diffusing photons, regulates the luminosity and determines the observed spectrum.

In \cite{Svirski+2012} we discussed the observed signal from events with breakout times of a few days after the SN explosion or longer.
For standard SNe parameters these events map to a breakout shock velocity $v_{bo,9}\le1$, in units of
$10^9 {\rm~cm~s^{-1}}$.
The analysis in \cite{Svirski+2012} is limited to scenarios in which the radiation emitted by the cold medium that surrounds the hot layer is in thermal equilibrium with the emitting electrons.
Svirski et al. (in preparation) discuss non-equilibrium scenarios.

Here we turn  our focus to a yet  unexplored regime of interacting  SNe, that of
extremely high  shock velocities  found in  compact progenitor  explosions. WR
progenitor SNe are  expected to develop $v_{bo,9}\gtrsim  4$, corresponding to
breakout times of minutes rather than days (see Equation \ref{EQ v_bo} below).
In addition,  WR stars  go through a massive  mass loss and  thus  some are likely to
explode through a  thick wind. We find  that the emission from  such events is
X-ray  dominated, and  therefore their  detection is  challenging both  due to
their short duration and due to the  narrow field of view that current sensitive X-ray
detectors  provide. However,  the serendipitous  X-ray detection  of SN  2008D
\citep{Soderberg+2008},  a  SN of  an  extreme  shock velocity  that  possibly
exploded through a  thick wind, encourages us to study  the parameter space of
such                                                                   events.

If the progenitor of a core-collapse SN  is \emph{not} surrounded by a wind, then the early SN luminosity is powered by the internal energy that the SN shock deposits in the progenitor's envelope. If, in contrast, a thick wind \emph{does} surround the progenitor, then the interaction of the expanding ejecta with the wind dominates the event luminosity (at least as long as the unshocked wind is optically thick). In this case, the radiation deposited in the envelope by the shock, rather than escaping freely to the observer, diffuses further through the interaction layer and affects its cooling. Although the energy flux of this radiation may be considerably lower than the flux powered by the interaction, we find that it has a profound influence on the observed luminosity and spectrum of a WR exploding through a thick wind.
While a failure to account for this radiation results in a faint monochromatic signal of $T \gtrsim 60$ keV, its proper inclusion invokes an efficient cooling of the shocked plasma by inverse Compton interactions and implies a much higher luminosity -- comparable to the shock kinetic energy flux, and a rather flat, $\nu F_{\nu}= Const$, observed spectrum that spans over a large frequency range that ends at $T<60$ keV.

Our analysis applies to SNe that explode through a thick wind and reach a breakout velocity $v_{bo,9}\gtrsim 3$,
including highly energetic explosions in blue supergiant and red supergiant progenitors, in cases that they are surrounded by a thick wind.
While motivated by SNe that explode through a thick stellar wind, the solution that we provide is general and may apply to other physical settings that involve a fast collisionless shock in a thick medium.

In Section \ref{sec:analysis} we develop a general solution for the spectrum of a fast collisionless shock propagating through an optically thick medium, in the presence of an external soft radiation source. In Section \ref{sec:impl} we review the implications of our analysis for fast shock SNe exploding through a thick wind. In Section \ref{sec:post} we discuss the evolution of such systems as the wind ahead of the shock becomes optically thin.
We summarize in Section \ref{sec:conc}.

\section{An irradiated fast collisionless shock in a thick medium}\label{sec:analysis}

\subsection{Essential hydrodynamics}\label{sec:hydro}

The hydrodynamics of the interaction with a wind is discussed in \cite{Chevalier_Irwin2012} and
\cite{Svirski+2012}, based on the self-similar
solution of \cite{Chevalier1982}. We assume here a spherical shock, but our analysis should remain valid under a mild asphericity. For a standard wind density profile, $\rho_w\propto r^{-2}$, and a compact progenitor (a density profile with a polytropic index of $3$ near the stellar edge), the shock radius at early stages of the interaction evolves
with time as $r(t)\propto t^{0.875}$
and its velocity decreases slowly, $v_s(t)\propto t^{-0.125}$.

We refer to a shock as a \emph{fast cooling} one (not to be confused with the shock velocity) if the shock is cooled within a dynamical time, i.e. by the time the shock radius doubles. A longer cooling time defines a \emph{slow cooling} shock.
Under a fast cooling, the bolometric luminosity at and following the breakout is \citep{Svirski+2012}\footnote{Equations \ref{eq:L(t)} and \ref{EQ v_bo} presented here slightly differ from \cite{Svirski+2012}, since a density profile with a polytropic index of $3$ is assumed, as expected in a Wolf-Rayet.}:
\small
\begin{equation}\label{eq:L(t)}
L_{fast}(t)\sim 3.5 \times 10^{43}\,t_{bo,m}\,\left(\frac{v_{bo}}{10^{10} {\rm~cm~s^{-1}}}\right)^3 \left(\frac{t}{t_{bo}}\right)^{-0.375}\,\rm{\frac{erg}{s}}
\end{equation}
\normalsize
where $t_{bo}$ is the breakout
time and $t_{bo,m}=t_{bo}/{\rm minute}$. Observationally $t_{bo}$ is
also roughly the rise time of the breakout pulse.

The shock velocity at the breakout depends on the explosion energy, $E$, the mass of the SN ejecta, $M_{ej}$, and the breakout time \citep{Svirski+2012}:
\begin{equation}\label{EQ v_bo}
    v_{bo} \approx 6\times 10^9\,  M_{5}^{-0.31} E_{51}^{0.44}t_{bo,m}^{-0.25}{\rm~\frac{cm}{s}~}
\end{equation}
where $E_{51}=E/10^{51} {\rm~ erg~s^{-1}}$ and $M_{5}=M_{ej}/5M_\odot$.

\subsection{The shock cooling channels}\label{sec:system}

The collisionless shock that forms after the breakout produces a layer of hot shocked electrons, of a temperature $T_h \ge 60$ keV, immediately behind the shock \citep{Katz+2011,Murase+2011}. This hot layer is the main energy source for the observed radiation, but photons leaving it must diffuse through the adjacent colder layers before escaping to the observer.

\begin{figure}
	\centering
		\epsscale{1.2} \plotone{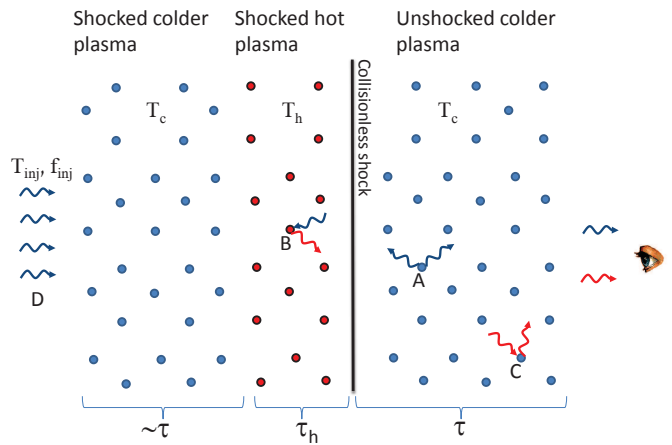} 
	\caption{A schematic illustration of a collisionless
shock propagating in an optically thick medium.}
	\label{fig:layers}
\end{figure}

Figure \ref{fig:layers} illustrates a fast cooling collisionless shock propagating through an optically thick medium. Electrons in the unshocked layer emit soft photons, of a temperature $T_c$ comparable to their own (A in Figure \ref{fig:layers}). Some of these soft photons diffuse through the layer of hot shocked electrons and cool them through IC scattering, thus gaining energy (B in Figure \ref{fig:layers}). The scattered photons then diffuse through the unshocked electrons and heat them by Compton scattering (C in Figure \ref{fig:layers}), causing them to emit photons once again, as in step A, thus completing a self-sustaining cycle of IC cooling.

Before it escapes, the radiation energy produced by the cooling shocked gas diffuses not only through the unshocked, upstream layer, but also through a layer of cooled shocked gas behind the hot shocked layer, of an optical depth similar to that of the upstream, $\sim\tau$ (leftmost layer in Figure \ref{fig:layers}). Exposed to the same photon population, this cold shocked layer stops cooling at a temperature comparable to that of the upstream (assuming its cooling is IC dominated, see discussion in Appendix \ref{sec:compression}). The total optical depth of the cold layer is thus $\sim 2\tau$.

A second source for IC cooling is the interaction of hot shocked electrons with \emph{external} soft photons, injected into the interaction region after being produced elsewhere, e.g. the diffusion of photons produced by a radiation mediated shock prior to its breakout (D in Figure \ref{fig:layers}).
This radiation is characterized by its temperature, $T_{inj}$, and its luminosity, which we parameterize by $f_{inj}$, the ratio between this luminosity and the energy produced by the shock, $L_{fast}$.
In addition to IC cooling, the hot layer also cools through free-free emission of hard photons. Such photons transfer most of their energy to the cold layer electrons through Compton scattering before they escape.

\subsection{The physical picture in a nutshell}\label{sec:picture}

The energy source for the observed radiation is the hot shocked layer. In a fast shock, $v>30,000$ km/s, this layer cools efficiently only if exposed to an external soft radiation which photon number flux is above some threshold. Then, most of the energy in the system is carried by the external photons injected into the interaction region. If the number flux of these photons is only mildly above the efficient cooling threshold, they cannot gain the required energy directly from scattering with the hot layer, because the hot layer is optically thin and the respective IC heating rate at the photon's injection energy is too low. In this case, the mediating component, which delivers the energy from the hot layer to the soft injected photons, is the surrounding colder layer. The mediation is realized as follows: Photons already heated above the injection energy go through further IC upscattering by the hot layer, and subsequently heat the colder layer through Compton scattering. The heated colder layer, which is optically thick, quickly upscatters newly injected soft photons to the required energy.
Below we derive the efficient cooling threshold, and find the radiation spectrum that develops and supports a steady state mediation process. The spectral slope and range depend on the injected photons number flux, and the solution we provide accounts also for a flux high above the threshold, where photons gain energy mainly from direct scattering by the thin hot layer.

\subsection{The cooling of an isolated fast shock}\label{sec:threshold}

Let us first consider a shock that is isolated from external radiation sources, such that the cold layer electrons are the only source for production of soft photons that interact with the hot shocked layer.
If the shock is cooling fast, the cold layer electrons are heated by a radiation energy flux that is determined by the shock velocity and scales as $v^3$ (Equation \ref{eq:L(t)}).
The heated cold layer electrons cool by free-free and bound-free emission, and their temperature is determined by the balance between their Compton heating rate and free-free/bound-free cooling rate.

Since the free-free/bound-free emissivity of thermalized electrons is proportional to $\sqrt{T}$, and the heating scales as $v^3$, the cold layer balance temperature, $T_c$, is extremely sensitive to the shock velocity: $T_c\propto v^6$.
Thus, a small increase of the shock velocity implies a large rise of $T_c$, such that above some threshold shock velocity, emission processes are no longer able to cool the cold layer electrons below the temperature of the hot shocked layer, $T_h$. In Appendix \ref{sec:fast cooling} we find this threshold shock velocity to be $\sim 30,000$ km/s.

A velocity above the threshold implies $T_c\sim T_h$ and as a result, photons emitted by cold layer electrons (A in Figure \ref{fig:layers}) are too hard to cool the hot layer\footnote{Photons from the softer free-free tail cannot cool the hot layer, see Appendix \ref{sec:fast cooling}.}. IC cooling is then ruled out and the hot layer electrons are subjected to a free-free cooling. However, at $v\gtrsim 20,000$ km/s free-free cooling of the hot layer becomes slow (Appendix \ref{sec:fast cooling}). Therefore, if the shock is faster than the threshold velocity, and devoid of any external radiation source, its cooling is slow and we expect an observed radiation temperature $T_{\gamma}\sim T_h \ge 60$ keV and an observed luminosity lower than implied by Equation \ref{eq:L(t)}.

\subsection{The cooling of a fast shock in the presence of an external radiation}\label{sec:conditions}

Consider now a fast ($v\gtrsim 30,000$ km/s) shock that is exposed to an external flux of soft photons, with a characteristic injection temperature $T_{inj}\ll T_h$. Soft photons invoke an IC cooling of the hot shocked layer. If the injected photon number flux is high enough then IC dominates the shock cooling, and above some flux threshold it also supports a fast cooling of the shock. We next find the criterion under which the injected radiation supports a fast cooling.

Recalling that a slow cooling implies a cold layer temperature $T_c\sim T_h$, consider a marginal external flux that is just enough to cool the shock within a dynamical time, thus marking the limit between slow and fast cooling. In this case, the cold layer temperature remains at $T_h$, as in a non-marginal slow cooling. Let us further assume that the Compton y-parameter of the entire interaction region, y, is large enough to upscatter an injected photon up to $T_h$ before it escapes to the observer. In such a case, if the number flux of injected soft photons (the number of photons injected into the interaction region per second), $\mathcal{N}_{inj}$, is sufficient to carry, after photons are upscattered to $T_h$, the luminosity of a fast cooling shock, $L_{fast}$,
then a fast IC cooling replaces the free-free dominated cooling picture described above (Section \ref{sec:threshold}).
The threshold conditions for a fast IC cooling by an external soft radiation source are therefore:

\begin{subequations}\label{eq:condition}
\begin{equation}
\mathcal{N}_{inj}=\frac{f_{inj}L_{fast}}{T_{inj}}>\frac{L_{fast}}{T_h},
\end{equation}
or
\begin{equation}
f_{inj}>\frac{T_{inj}}{T_h};
\end{equation}
\end{subequations}
where we assumed $T_{inj}e^y= T_h$, i.e.
\begin{equation}\label{eq:condition2}
y>\ln\left(\frac{T_h}{T_{inj}}\right),
\end{equation}
where $T_{inj}$ is the injection photon temperature and $f_{inj}$ is the ratio between the luminosity of the external source and $L_{fast}$.

Meeting the conditions of Equations \ref{eq:condition} and \ref{eq:condition2} has two major implications:
(I) Under a fast cooling, the bolometric luminosity is simply the kinetic energy flux through the shock, given by Equation \ref{eq:L(t)}. (II) The injected photon number flux is conserved, since compared to it, free-free/bound-free emission is negligible and photons are scarcely added to the system. Photon absorption is also negligible since $T_c$ is still much higher than the cold layer blackbody temperature, $\lesssim 50$ eV \citep{Svirski+2012}. Therefore, the photons that reach the observer are the very same photons injected into the system, and in order to find the observed spectrum we have to follow the interaction of these photons as they diffuse through the hot shocked layer and the cold layer that surrounds it.

\subsection{Photon energy gain from the hot shocked layer}

Let us first examine the photons gain from interactions with the hot shocked layer alone.
With heating and cooling balanced, the cold layer electrons do not add or subtract \emph{net} energy from the diffusing radiation. Therefore, the energy density of the diffusing radiation is set by the hot shocked layer alone.
A fast cooling implies a diffusing radiation with an energy density $\epsilon_{rad}\sim\epsilon_{gas}\frac{t_{diff}}{t}$ in the vicinity of the shock, where $t_{diff}$ is the diffusion time across the cold medium, $t$ is the dynamical time and $\epsilon_{gas}\sim \rho v^2$ is the energy density of the shocked gas. Since after the breakout $t_{diff}<t$, the system assumes a steady state and the radiation energy density is rather constant across a dynamical time.

A constant radiation energy density implies that each energy contribution by the hot shocked layer, through photon upscattering, is matched by an escape of a radiation carrying the same energy. Since a photon's probability to hit an electron as it crosses the hot shocked layer and its probability to escape the system before the next crossing of the hot layer are both independent of its energy (for Newtonian shocks, where $\sigma_T$ applies), a photon's average fractional gain as it crosses the hot shocked layer must be equal to its probability to escape the system before the next crossing, regardless of its frequency.

A steady state implies $y_h=1$, where $y_h$ is the Compton y-parameter of the hot layer alone, given by the average fractional gain per passage through the hot layer, times the average number of passages -- which is one over the escape probability. Since a steady state implies an average fractional gain equal to the escape probability, it also imposes a $y_h=1$\footnote{Section \ref{sec:conditions} describes a scenario in which the cold layer dominates the photon gain, i.e. $y_c\gg 1$ and hence $y=y_h+y_c\gg 1$.}. In addition, since a single photon scattering by the hot layer already satisfies a $y\sim 1$ (as $\frac{4T_h}{m_ec^2}\sim 1$), the multiple passages through the hot shocked layer implied by a cold layer of $\tau\gg 1$ impose an optically thin hot layer, $\tau_h\sim 1/\tau$.

A $y_h=1$ implies that a typical photon doubles its energy due to IC scattering with the hot layer before it escapes. However, a small fraction of the photons goes through repeated IC scattering with the hot layer and reaches higher energies. As in a Fermi acceleration, the process of repeated scattering with constant fractional energy gain and escape probability generates a power-law spectrum, which slope is determined by the ratio between the fractional gain and the escape probability. If the two are equal, a flat $\nu F_{\nu}= Const$ spectrum develops, carrying a constant energy per logarithmic frequency scale. The emergence of such flat sepctrum is further explained in Svirski et al. (in preparation).

\subsection{Photon energy gain from the cold layer}

Photons injected into the interaction region are scattered both by the hot thin layer and the colder thick layer. If photons only gain energy from scattering by the hot layer, then the spectrum of the diffusing radiation is flat, as discussed above. However, if the photons injection temperature is lower than the thick layer balance temperature $T_c$, then the photons also gain energy from scattering by the cold thick layer. In this case $y=y_h+y_c>1$, where $y_c>0$ refers to gain from the cold layer. As we show in section \ref{sec:impl}, SNe that explode through a thick wind indeed imply $T_{inj}\ll T_c$, and therefore we should also consider the gain from scattering by cold layer electrons.

The threshold scenario presented in Section \ref{sec:conditions} represents an extreme photon gain from cold layer electrons, where \emph{every} injected photon is gradually upscattered to the same final temperature $T_h$. This implies, under a constant injection of soft photons and a scattering probability that is independent of the photon frequency, a $\nu F_{\nu}\propto \nu$ spectrum, where each logarithmic scale carries an equal \emph{number} of photons rather than an equal \emph{energy}.

\begin{figure*}
	\centering
    \subfigure[]
    {
        \centering
        \epsscale{1.1} \plotone{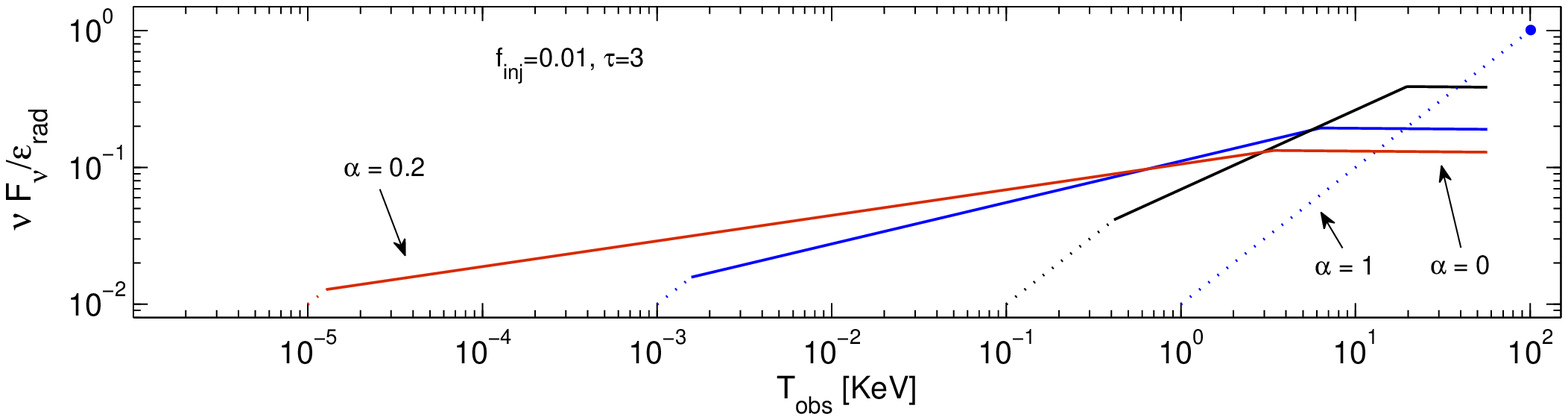}
        \label{fig:spectra1}
    }
    \subfigure[]
    {
        \centering
        \epsscale{1.1} \plotone{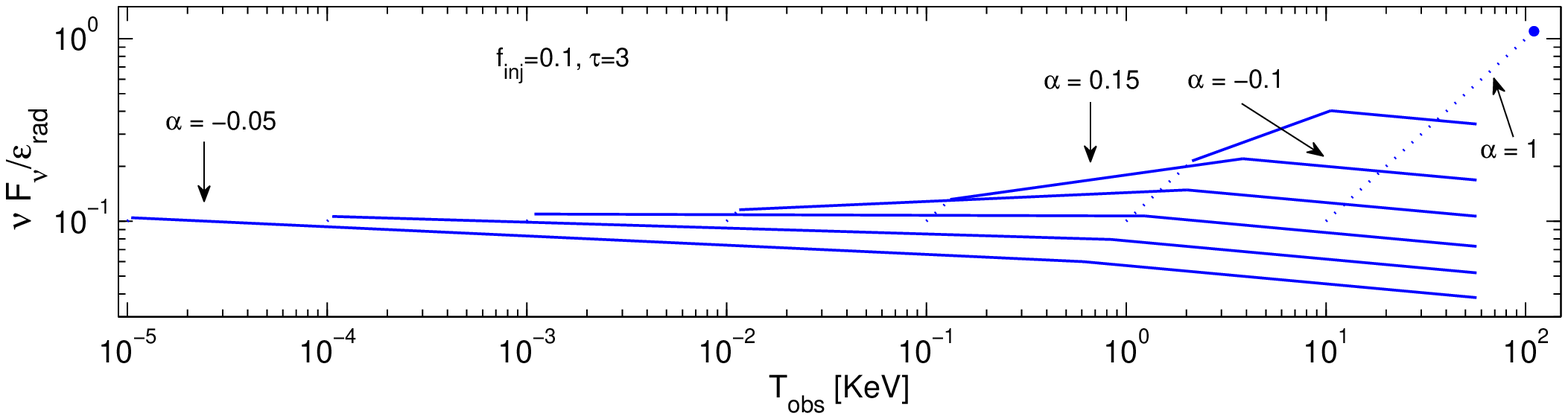}
        \label{fig:spectra2}
    }
    \caption{Spectral slope and range for various photon injection temperatures, $T_{inj}$ (the left dotted end of each curve), corresponding to various injected photon number flux, for $\tau=3$, typical of an interacting Wolf-Rayet SN. The energy flux of injected photons as a fraction of the shock flux is fixed, $f_{inj}=0.01$ in (a) and $f_{inj}=0.1$ in (b). Spectral slopes are rather flat for lower $T_{inj}$ values that we consider vs. free-free like for the higher values. Per given $T_{inj}$, the spectral slope in (b) is flatter than in (a), due to the higher $f_{inj}$. Wolf-Rayet progenitors that explode through a thick wind have $f_{inj}\sim 0.1$ and $T_{inj}\sim 10-100$ eV, thus producing a rather flat observed spectrum (see text). The dotted part of each spectrum is present in the diffusing radiation but is not observed since a typical photon is upscattered to the solid part of the curve before escaping to the observer. In spectral slopes $\alpha \rightarrow 1$, i.e. the rightmost curve in (a) and (b), frequencies below the upper cutoff do not reach the observer, and the observed spectrum is monochromatic.}
    \label{fig:spectra}
\end{figure*}

Therefore, we expect the spectral slope $\alpha$ in $\nu F_{\nu}\propto \nu^{\alpha}$ to vary, according to the injected photon number flux, between $1$, when the flux is low (marginally satisfying Equation \ref{eq:condition}) and the cold layer dominates the gain, and $0$, when the flux is high and the hot thin layer dominates the gain. Figure \ref{fig:spectra1}, in which each line represents a spectrum produced by a different injected photon number flux, demonstrates the flattening of the spectral slope as we increase the injected flux. The figure is based on Equations \ref{eq:energy_balance}-\ref{eq:Tmin} below.

How many injected photons are enough to produce a flat spectrum? To boost photons, the cold layer electrons first need to gain energy. A photon diffusing through a cold medium of an optical depth $\tau$ suffers significant Compton loses only if its energy is above $m_ec^2/\tau^2$. Therefore, if an average photon energy gain $\ll m_ec^2/\tau^2$ is sufficient to carry the shock energy, i.e.
\begin{equation}\label{eq:flat1}
\mathcal{N}_{inj}\gg \frac{L_{fast}}{m_ec^2/\tau^2}
\end{equation}
or
\begin{equation}\label{eq:flat2}
f_{inj}\gg \frac{T_{inj}}{m_ec^2/\tau^2},
\end{equation}
then photons do not lose energy by scattering with cold layer electrons, hence the electrons remain cold, and a flat spectrum evolves. In this case $y=y_h+y_c\approx y_h\approx 1$.

In contrast, if the injected photon number flux is low, requiring each photon to reach an energy $\gg m_ec^2/\tau^2$ in order to jointly carry the shock energy, then most of the energy that photons gain from the hot layer is transferred to cold layer electrons.
Since a steady state implies a fixed $T_c$, the energy transferred to the cold layer electrons is further delivered to the diffusing photons, such that the cold layer contributes most of the photon gain, $y=y_h+y_c\gg 1$, producing a non-flat spectrum. Thus, the condition
\begin{equation}\label{eq:ff1}
\frac{T_{inj}}{T_h}<f_{inj}\ll \frac{T_{inj}}{m_ec^2/\tau^2}
\end{equation}
implies a $\nu F_{\nu}\propto \nu$ spectrum of the radiation diffusing within the interaction layer, although an observer will only see the upper cutoff of this spectrum for the following reason. The observed spectrum in this case is different from the diffusing spectrum because photons are injected at $T_{inj}$ but typically upscattered to $T_{inj}e^{y_c}$ before escaping. Therefore, while the spectrum of the \emph{diffusing} radiation always starts from $T_{inj}$, the \emph{observed} spectrum starts at a higher temperature, $T_{min}=T_{inj}e^{y_c}$. A flat spectrum, of $y_c< 1$, implies $T_{min}\sim T_{inj}$ and an observed spectral range $T_{inj}\lesssim T<T_{max}$. However, if $y_c\gg 1$, a typical photon is upscattered to the spectrum's upper cutoff $T_{max}$ before it escapes, and a monochromatic radiation is observed.

The spectrum upper cutoff depends on the slope: In a flat spectrum, the cold layer electrons are heated by very few hard photons and as a result their temperature is kept low, $T_c<m_ec^2/\tau^2$. The hardest photons reach $T_{max}\sim m_ec^2/\tau^2$, beyond which they lose their excess energy in Compton scattering. When less photons are injected and the spectrum is harder, more energy is carried by hard photons and $T_c$ gradually rises above $m_ec^2/\tau^2$. Then, $T_{max}\sim T_c$ since photons below $T_c$ do not lose energy by Compton scattering.

Hence, equation \ref{eq:ff1} implies an observed monochromatic signal with a photon temperature $T_{max}\sim T_c\sim\frac{T_{inj}}{f_{inj}}$, reflecting an equipartition of the shock energy among the injected photons. The rightmost curve in Figure \ref{fig:spectra1}, where the dotted part is not observed, depicts such a spectrum. At the other extreme, Equation \ref{eq:flat2} implies a rather flat observed spectrum spanning across $T_{inj}\lesssim T\lesssim m_ec^2/\tau^2$. Note that if $T_c<m_ec^2/\tau^2$, the spectrum in the the range $T_c\lesssim T\lesssim m_ec^2/\tau^2$ is exactly flat since cold layer electrons do not contribute energy to photons above their temperature\footnote{\label{foot:4T}In fact the spectrum break is at $\approx 4T_c$ rather than $T_c$, since photons below $4T_c$ still gain energy from $T_c$ electrons. For simplicity we ignore this factor here but include it in Equations \ref{eq:energy_balance}-\ref{eq:slopes2} below.}. The spectra in Figure \ref{fig:spectra1} (excluding the rightmost) exemplify the flat section in the range $T_c< T< m_ec^2/\tau^2$.


\subsection{A derivation of the spectrum}

We now derive the spectral slope $\alpha$ and the cold layer temperature $T_c$ for a given external radiation with $T_{inj}$ and $f_{inj}$, and a given optical depth $\tau$.
The spectral slope is determined by the ratio between a photon's average fractional energy gain $\frac{\left\langle \Delta e_\gamma\right \rangle}{e_\gamma}$ over some time period (e.g. subsequent crosses of the hot thin layer) and its probability to escape during this period, $P_{esc}$. As in a Fermi acceleration, the spectral slope $\alpha$ in $\nu F_{\nu}\propto \nu^{\alpha}$ is
\begin{equation}\label{eq:fermi}
\alpha=1-P_{esc}/\frac{\left\langle \Delta e_\gamma\right \rangle}{e_\gamma}.
\end{equation}
Here we have to account for the gain from both the hot and cold layers.
Since the $y$ parameter within a given layer is $\left\langle \Delta e_\gamma\right \rangle$ over the time that a typical photon escapes the system, we can use
$y_h=1$ and $y_c=(2\tau)^2\frac{4T_c}{m_ec^2}$ (the cold layer's optical depth is $2\tau$) to obtain
\begin{equation}
\alpha=1-\left[1+(2\tau)^2\frac{4T_c}{m_ec^2}\right]^{-1}=1-\frac{1}{y}.
\end{equation}
The dependence of the spectral slope on $T_c$ confirms the two regimes we discussed earlier: $T_c\gg m_ec^2/\tau^2$ implies $\alpha\approx1$ while $T_c\ll m_ec^2/\tau^2$ implies $\alpha\approx0$.

In order to simplify the discussion we have thus far ignored a further effect of the external radiation. The addition of an external radiation field results in a total radiation energy density that is higher than the one implied by the fast cooling shock alone. This enhancement implies a photon escape probability that is higher than the average gain from the hot layer, because the energy that escapes is larger then the energy of the cooling shocked gas. For example, if $f_{inj}=1$, i.e. an external radiation energy flux equal to that of the cooling shock, then the gain from the shock supplies only half of the escaping energy flux. Therefore, in the general case both $y_h$ and $y_c$ are reduced by a factor $(1+f_{inj})^{-1}$. For the hot shocked layer this effect is physically manifested through a decrease of $\tau_h$ by this factor, whereas in the cold layer $T_c$ is reduced by the same factor. The reduced gain implies softer spectral slopes. Specifically, the flat spectrum in the range $T_c<T<m_ec^2/\tau^2$, if exists, is replaced by $\nu F_{\nu}\propto \nu^{-f_{inj}}$.

Figure \ref{fig:spectra1} depicts spectra that match $f_{inj}=0.01$, where the correction factor is negligible, while Figure \ref{fig:spectra2} demonstrates the reduced slopes implied by a larger (and likely realistic for WR SNe) $f_{inj}=0.1$. Note the slightly negative slopes at the left side of Figure \ref{fig:spectra2}, and the $\alpha=-f_{inj}$ that replaces the flat spectrum in the range $T_c<T<m_ec^2/\tau^2$. As expected, if $f_{inj}\gg1$ then the spectrum reduces to the photon injection temperature $T_{inj}$.

Accounting for the various effects described above, we now formulate the equations for solving the balance temperature of the cold layer electrons, $T_c$, and the spectral slope in the range\footnote{See footnote \ref{foot:4T}} $T_{inj}<T<4T_c$.
We denote the spectral index up to $4T_c$ by $\alpha_1$, as in $\nu F_{\nu}\propto\nu^{\alpha_1}$. The respective spectral index above $4T_c$ (due to gain from the
hot thin layer alone), if exists, is $-f_{inj}$, and although it is known we express it, for clarity, as $\alpha_2$.

The first equation expresses the energy balance:
\begin{equation}\label{eq:energy_balance}
\int^{4T_c}_{T_{min}}L_{\nu}(\alpha_1)d\nu +\int^{T_{max}}_{4T_c}L_{\nu}(\alpha_2)d\nu=(1+f_{inj})L_{fast}.
\end{equation}
The second equation expresses $\alpha_1$ as a function of $T_c$:
\begin{subequations}\label{eq:slopes2}
\begin{equation}
\alpha_1=1-\frac{1}{y_h+y_c},
\end{equation}
where
\begin{equation}
y_h=\frac{1}{1+f_{inj}}\,,\,\,\,\,y_c=(2\tau)^2\frac{4T_c}{m_ec^2}.
\end{equation}
\end{subequations}
The range of the spectrum is:
\begin{subequations}\label{eq:Tmin}
\begin{equation}
T_{min}=\min\left\{T_{inj}e^{y_c},4T_c\right\},
\end{equation}
and
\begin{equation}
T_{max}=\max\left\{\frac{m_e c^2}{\tau^2},4T_c\right\}.
\end{equation}
\end{subequations}
Solutions for various $f_{inj}$ and $T_{inj}$ values are presented in Figure \ref{fig:spectra}. The dotted part of each spectrum depicted in the figure indicates the injection temperature, but is not observed, since a typical photon reaches $T_{min}$ before it escapes. The slope of the unobserved part is $\alpha=1$ since $P_{esc}\rightarrow 0$.

\section{Implications for fast SN shocks propagating through a thick wind}\label{sec:impl}

\subsection{The observed signal}

If the optical depth of the wind surrounding a star is $>c/v$, the first photons escape once the optical depth of the unshocked wind reaches $\sim c/v$, producing a breakout pulse of a luminosity that follows Equation \ref{eq:L(t)}. For WR progenitors, this pulse has a non-thermal spectrum that peaks at a few keV \citep[e.g.][]{Nakar_Sari2010,Sapir+2011}.
Once the radiation mediated shock breaks out, photons cannot mediate the shock further, and a collisionless shock is formed within a dynamical timescale. The time integrated X-ray signal is dominated by the radiation from the cooling collisionless shock, which luminosity also follows Equation \ref{eq:L(t)}, and which spectrum we now discuss.

When a SN ejecta expands into a thick wind, its interaction with the wind is the main energy source for the bolometric luminosity, but it is not the only source of radiation within the interaction layer. Prior to the interaction with the wind, the shock traverses the progenitor's envelope, unbinds it, and leaves behind a radiation that escapes at later stages. This radiation constitutes the early SN emission when there is no thick wind. When thick wind exists it plays the role of a sub-dominant soft radiation source, external with respect to the interaction region. Below we estimate the $f_{inj}$ and $T_{inj}$ values that this radiation implies for SNe that explode through a thick wind, and apply our model to find the expected observed signature.

The temperature and energy flux of the external soft radiation evolve as in a standard SN (one with no wind), because this radiation is released by an ejecta layer that is deeper than the reverse shock front and it is thus independent of the interaction with the wind. This is evident by comparing the mass of the ejecta layer releasing the soft radiation to the ejecta mass swept by the reverse shock.
At the breakout, photons escape from the wind layer and ejecta layer shocked by the interaction with the wind. The diffusion time from ejecta layers not yet shocked is longer than the dynamical time and therefore the photons in them are trapped. The diffusion time from the ejecta layer swept by the reverse shock is comparable to the dynamical time and therefore, it releases both the radiation energy produced by the reverse shock, and the radiation energy deposited by the earlier traverse of the forward unbinding shock through this envelope layer. After the breakout, the ejecta shell that satisfies $\tau=c/v$, from where soft photons deposited by the unbinding shock can diffuse to the interaction layer (these photons dominate the SN light when there is no wind), recedes inwards faster than the reverse shock front. The ejecta mass swept by the reverse shock scales, like the wind mass swept by the forward shock, as $m\propto r\propto t^{0.875}$, while photons can diffuse to the interaction layer from a shell (noted as the luminosity shell in \citealt{Nakar_Sari2010}) which mass grows faster, as $m\propto t^{1.75}$.

The velocity that the unbinding shock had when it crossed the ejecta layer that is included in the interaction layer at the breakout time is comparable to $v_{bo}$, and the mass of the shocked ejecta is comparable to the mass of the shocked wind. However, the energy deposited by the unbinding shock suffered adiabatic loses during the expansion. As a result,
during the breakout the energy of the injected soft radiation is lower than the interaction energy by the ratio:
\begin{equation}\label{eq:loses}
\frac{E_{inj}}{E_{int}}
\approx \left(\frac{R_{bo}^3}{R_*^2d_i}\right)^{-\frac{1}{3}}=\frac{R_*}{R_{bo}}\left(\frac{d_0}{R_*}\right)^{\frac{1}{3}}\left(\frac{d_i}{d_0}\right)^{\frac{1}{3}},
\end{equation}
where $R_{bo}$ is the breakout radius, $R_*$ is the stellar radius, and $d_i$ is the initial width (before expansion starts) of the breakout layer. We also introduce $d_0$, the initial width of the fastest moving ejecta layer before interaction with the wind start (this would have been the breakout layer if there were no wind). Since $v_{bo}\sim v_0$, where $v_0$ is the shock velocity as it crosses the stellar edge, we estimate $\left(\frac{d_i}{d_0}\right)^{\frac{1}{3}}=\left(\frac{v_{bo}}{v_0}\right)^{\frac{1}{0.57n}}\sim 1$ ($1.5<n<3$ is the power-law index of the pre-explosion stellar density
profile near the edge) and by approximating $\left(\frac{d_0}{R_*}\right)^{\frac{1}{3}} \approx \frac{1}{5}$ (\citealt{Nakar_Sari2010}), we obtain an estimate of $f_{inj}$ at the breakout
\begin{equation}\label{eq:ratio}
f_{inj}=\frac{E_{inj}}{E_{int}}\approx \frac{1}{5} \frac{R_*}{R_{bo}}.
\end{equation}

WR progenitors are typically surrounded
by a wind of an optical depth $\tau_w\lesssim 20$ \citep{Crowther+2007}, and have
a breakout optical depth $\tau_{bo}\approx c/v\sim 5$, implying $R_{bo}\lesssim 4R_*$ (since $r\propto 1/\tau$), hence:
\begin{equation}\label{eq:ratio_wr}
f_{inj}\sim 0.1\,\,(\rm{WR}).
\end{equation}

Following breakout, the interaction luminosity evolves as $t^{-0.37}$ (Equation \ref{eq:L(t)}), whereas the luminosity of the injected soft radiation evolves as $t^{-\alpha}$, with $0.17\le\alpha\le0.35$ depending on the respective $1.5<n<3$ value \citep{Nakar_Sari2010}. Thus, $f_{inj}$ is rather constant and Equation \ref{eq:ratio_wr} remains valid also after the breakout. The injected radiation temperature $T_{inj}$ for various core-collapse SN progenitors is given by \cite{Nakar_Sari2010}. It typically softens with time, within the range
\begin{equation}\label{eq:T_inj}
1\lesssim T_{inj}\lesssim 100 \,\rm{ev}.
\end{equation}

Equations \ref{eq:ratio_wr} and \ref{eq:T_inj} imply that a WR exploding through a thick wind satisfies both the fast cooling threshold condition (Equation \ref{eq:condition}) and the condition for a flat spectrum (Equation \ref{eq:flat2}). Its bolometric luminosity thus follows Equation \ref{eq:L(t)} and it has a flat $\nu F_{\nu}= Const$ spectrum across an initial frequency range $0.1\lesssim T\lesssim 50$ keV. As $\tau$ and $T_{inj}$ decrease with time the range of the flat spectrum becomes wider with time up to $0.01\lesssim T\lesssim 100$ keV before the wind becomes optically thin and the X-ray signal fades away. Note that the flat spectrum is independent of the (Newtonian) shock velocity, the metallicity, the hot layer temperature, and the wind density profile.

Since WR progenitors may eject thick winds during their WR phase, we expect the scenario of exploding through a thick wind to be rather common among WR SNe. In Svirski \& Nakar (in preparation) we discuss the application of the model presented here to the observations of SN 2008D \citep{Soderberg+2008}, a type Ib/c SN which progenitor was likely a WR star.

Typically, more extended progenitors like red and blue supergiants, are not surrounded by a thick wind. However, there seem to be cases where SN explosions of such extended progenitors do take place within a thick wind. Our analysis applies to such cases only if they attain $v_{bo,9}>3$ (which requires, for red supergiants, a very large explosion energy) and satisfy Equation \ref{eq:condition}, which requires, for $T_{inj}> 1$ eV and $v_{bo,9}\sim 3$, a ratio $\frac{R_*}{R_{bo}}>0.02$. If the latter ratio approaches unity, the event will look similar to an exploding WR, while a ratio much lower will tend towards a monochromatic $50-100$ keV signal.

Finally, the soft SN radiation that dominates the cooling of fast shocks is present in slower shock SNe as well (as discussed, e.g., in \citealt{Svirski+2012}). However, the role it plays in such events is less pronounced, as the IC cooling is then dominated by photons emitted within the interaction later by the cold layer electrons.

\subsection{Some refining remarks}

Presenting our model we have made some implicit simplifications that deserve a discussion:
\begin{enumerate}
  \item We derived a fast cooling threshold condition (Equation \ref{eq:condition}) that depends on a minimum required $y$ parameter (Equation \ref{eq:condition2}). As the shock propagates towards the optically thin part of the wind, i.e. $\tau\rightarrow1$, the contribution of the cold layer to the photon gain diminishes and $y=y_h+y_c\approx 1$, implying a flat spectrum with a lower cutoff at $T_{inj}$. The spectrum's upper cutoff is then $T_h$ rather than $m_ec^2/\tau^2$, and the condition for a fast cooling is
      \begin{equation}
      f_{inj}\ge 1/\ln\left(\frac{T_h}{T_{inj}}\right).
      \end{equation}
For a typical WR, a fast cooling at $\tau=1$ requires $f_{inj}\gtrsim 0.1$, which is typically satisfied (Equation \ref{eq:ratio_wr}). At $\tau<1$ the cooling of fast shocks is always slow (see Appendix \ref{sec:fast cooling}) and our analysis breaks, as discussed in Section \ref{sec:post}.
  \item We ignored the radiation contribution by the reverse shock. The reverse shock carries a small fraction of the interaction energy, $\sim 1/30$ in a WR SN \citep{Balberg_Loab2011}. It transforms into a collisionless shock within half a dynamical time after the forward shock transition, and cools through similar processes. Due to its low energy content, the effect of the reverse shock radiation is minor, regardless of its temperature, unless $f_{inj}\lesssim 1/30$.
  \item The cooling of the post-shock gas from $T_p$ to $T_c$ implies, under a constant post-shock pressure, a significant compression of the gas. In Appendix \ref{sec:compression} we discuss the implications of such compression and show that it is unlikely to affect the signal predicted for a WR.
\end{enumerate}

\newpage

\section{The signal at $\tau<1$}\label{sec:post}

Once the optical depth of the unshocked wind drops below unity, fast cooling is no longer sustained\footnote{In fact, IC fast cooling may prevail until the total optical depth of the cold layer, including the cooled shocked layer, reaches unity, i.e. until $\tau\sim 1/2$. Since this is a small difference, we keep the notation $\tau<1$ to address the slow cooling phase.} (see Appendix \ref{sec:fast cooling}). The luminosity powered by the interaction then quickly drops and the soft radiation that characterizes a standard SN (with no wind) soon becomes the main source of bolometric luminosity. However, the interaction produces a faint X-ray signature that may still be seen.

At $\tau\ll1$, both IC and free-free cooling of fast shocks ($v_{bo,9}> 3$) are slow, and the interaction signal depends on the process that dominates the shock cooling.
Within a dynamical timescale from the transition to a slow cooling, the radiation field for IC cooling becomes the soft radiation alone, with no enhancement from the cooling shock. Then, the free-free vs. IC emissivity ratio is $\frac{\varepsilon^{ff}}{\varepsilon^{IC}}\sim\frac{\varepsilon^{ff,bo}}{\varepsilon^{IC,bo}}\frac{\tau_{bo}}{f_{inj}}\sim 0.3v_{bo,9}^{-3}f_{inj}^{-1}$ (using $\frac{\varepsilon^{ff,bo}}{\varepsilon^{IC,bo}}$ from \citealt{Svirski+2012} Equation 17) and it increases with time as $T^{-1/2}v^{-3}\propto t^{0.35}$. Accounting for the growth of $T_h$ that follows the breakout\footnote{Following breakout, the IC cooling rate of the hot shocked layer decreases due to the decrease of IC radiation field as the shock crosses the range $c/v\ge\tau\ge1$: $\epsr\propto \tau v/c\propto t^{-1}$. The decreasing cooling rate implies $T_h\approx60(t/t_{bo})^{0.4}$ keV \citep[][Equation 6, see also footnote \ref{foot:T_h} below]{Katz+2011}.} and for relativistic corrections yields
\begin{equation}\label{eq:post_ff_IC}
\frac{\varepsilon^{ff}}{\varepsilon^{IC}}(t)\approx 0.6v_{bo,9}^{-2.7}f_{inj}^{-1}\left(\frac{t}{t_{\tau=1}}\right)^{0.35}.
\end{equation}

\begin{figure}
	\centering
		\epsscale{1.2} \plotone{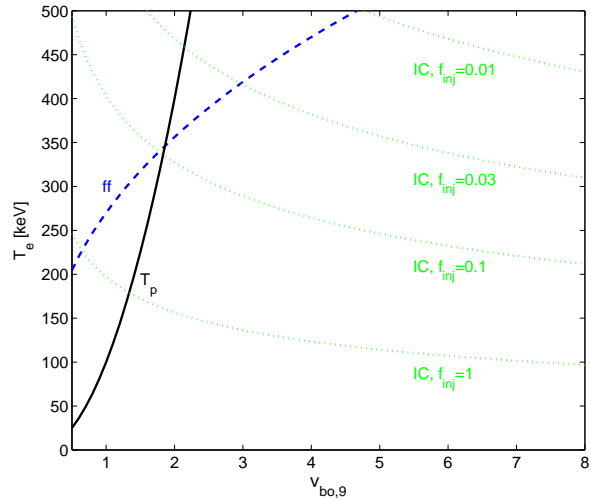}
	\caption{$T_h$ matching a balance between a heating of the shocked electrons by protons and a cooling by free-free emission (dashed line), IC interactions (dotted lines) and non, i.e. $T_h=T_p$ (solid line). IC cooling is sketched for different $f_{inj}$ values. The dominant cooling process corresponding to each shock velocity is the one yielding a lower $T_h$. High $v_{bo}$ and $f_{inj}$ imply an IC dominated cooling at $\tau\ll 1$.}
	\label{fig:post_diff}
\end{figure}

The dominant cooling process is the one that yields a lower balance temperature of the electrons in the hot shocked layer, $T_h$, assuming a heating by Coulomb collisions with shocked protons\footnote{\label{foot:T_h}$T_h$ under IC cooling is calculated as $T_h \approx 60\epsilon_{\gamma}^{-2/5}\gamma^{-4/5}$ keV (following \citealt{Katz+2011}), where $\gamma$ is the electrons Lorentz factor and $\epsilon_{\gamma}=\frac{v}{c}f_{inj}$ is the shock energy fraction available as a radiation field for IC cooling. $T_h$ under free-free cooling is calculated as $T_h\approx 270v_9^{0.4}$ keV, which approximates the balance temperature for Coulomb heating and free-free cooling when a relativistic correction for free-free emissivity is considered. At $\tau< 1$, $T_h$ depends only weakly on time, through $v$.}. Figure \ref{fig:post_diff} depicts $T_h$ at $\tau\ll 1$ as a function of shock velocity, under both free-free and IC cooling, with different $f_{inj}$ values considered. Since WR progenitors imply $v_9\gtrsim 4$ and $f_{inj}\sim0.1$, their cooling is dominated by IC, and so it remains due to the weak dependence on time (Equation \ref{eq:post_ff_IC}). The balance temperatures presented in Figure \ref{fig:post_diff} are rather constant in time, changing only through their dependence on the slowly decaying shock velocity.

We first discuss the regime of IC cooling dominance, relevant for WR SNe.
In a slow cooling, the complete layer of shocked wind remains hot. The shocked wind mass in a standard wind density profile is comparable to the unshocked wind mass that dominates the unshocked optical depth, and being swept forward by the ejecta, the shocked wind radius is comparable to the unshocked one, such that $\tau_h\sim \tau$.
The IC emissivity is dominated by single scattering of soft SN photons and therefore the interaction luminosity, relative to the dominant SN luminosity, evolves as $L_{int}/L_{SN}\propto \tau_h \propto\tau$. Substituting $L_{SN}\propto t^{-0.35}$ \citep{Nakar_Sari2010} then implies $L_{int,IC}\propto t^{-1.25}$.

Most of this energy is observed at a temperature $T_{obs}\sim T_{SN}\left(1+\gamma^2\frac{4T_h}{m_ec^2}\right)$, where $\gamma<2$ is the electrons Lorentz factor and $T_h<m_ec^2$, such that $T_{obs}$ is a few times $T_{SN}$.
However, a small fraction of the soft photons goes through multiple scattering and reaches higher temperatures.
At $\tau\ll 1$ the probability of a photon to go through $n$ collisions with the thin hot layer is $\tau_h^n$. Therefore, the luminosity observed at a given band $T>T_{SN}$ is attenuated by $\sim(\mu\tau)^n$, where $\mu=1+\gamma^2\frac{4T_h}{m_ec^2}$ is the single collision gain factor, and $n(T)=\ln(T/T_{SN})/\ln(\mu)$ is the number of collisions required to upscatter a $T_{SN}$ photon to a temperature $T$.
Such attenuation implies, at $\tau\ll 1$, a spectrum $\nu F_{\nu}\propto \nu^{\alpha}$ with $\alpha=1-\frac{\ln(1/\tau)}{\ln(\mu)}<0$, valid for $\tau\mu<1$
\footnote{This spectrum does not apply for $1>\tau>1/\mu$, because the $T_h$ value presented in Figure \ref{fig:post_diff}, and the implied $\mu$, are calculated for an IC emissivity degraded by a factor $f_{inj}$, but this degradation is completed only at $\tau\lesssim 1/\mu$. Over the range $1>\tau>1/\mu$, $T_h$ gradually grows towards its Figure \ref{fig:post_diff} value, continuously satisfying $\tau\mu(T_h)<1$. A higher $\mu$ would imply the formation of a hard spectrum and a recovery of fast cooling at $\tau<1$ (contradicting Appendix \ref{sec:fast cooling}). However, a fast IC cooling implies no $T_h$ rise, i.e. $\mu\sim 1$, and therefore $\mu>1/\tau$ is impossible.}.

During the time that $\tau$ drops from $\sim1$ to $\lesssim 1/\mu$, roughly a dynamical timescale, the X-ray luminosity has dropped by at least $f_{inj}$ of its fast cooling value (since the spectral slope is negative), and at $\tau< 1/\mu$ it decays as
\begin{equation}
L_{X,IC}(t)\propto\tau^{n_X}\propto t^{-n_X},
\end{equation}
where $n_X$ is the number of collisions that brings an Optical photon to the X-ray detector window. For a WR, where $T_h\sim 200$ keV (Figure \ref{fig:post_diff}), $3<n_X<4$ for \emph{Swift}/XRT and $5<n_X<6$ for \emph{NuSTAR}. Detecting such an X-ray luminosity decay pattern provides a unique evidence for an interaction of a WR SN with an optically thin wind.

Unlike typical WR SNe, interacting SNe with lower shock velocities, $v_{bo,9}\sim 3$, and lower injected energy flux, $f_{inj}<0.03$ (see Figure \ref{fig:post_diff}) cool, when $\tau\ll1$, by free-free emission rather than IC interactions. Following an initial drop by a factor $(0.6v_{bo,9}^{-2.7})^{-1}$ from the fast cooling luminosity to the free-free one (substituting $f_{inj}=1$ in Equation \ref{eq:post_ff_IC}), the interaction luminosity evolves as $\varepsilon_{ff}r^3 \propto n^2r^3 \propto r^{-1}$ where $\varepsilon_{ff}$ is the free-free emissivity, i.e.
\begin{equation}
L_{int,ff}(t)\propto t^{-0.9}.
\end{equation}
This emission is dominated by photons of a temperature comparable to that of the hot shocked electrons, $T_h\sim 400$ keV, such that the luminosity observed at a band $T$ is reduced, due to the free-free spectrum, by a factor $\sim T/T_h$ (e.g., by $\sim 80\rm{keV}$$/T_h$ in \emph{NuSTAR} upper band). Note that the luminosity in hard photons, $\gg 10$ keV, soon becomes dominated by free-free emission also for $v_{bo,9}> 3$.

The temperature depicted by the dashed line in Figure \ref{fig:post_diff} is an upper limit for $T_h$ and hence for the observed photons temperature under a free-free cooling dominance.
Free-free emission of $T_h$ photons implies pair production\footnote{Under IC dominance pair production is negligible, because a low fraction of electrons cooling via free-free emission of hard photons can only increase the overall electron/positron count by a low fraction.}, which decreases the temperature of the electrons and the emitted photons below the value in Figure \ref{fig:post_diff}, and increases the hot layer's optical depth. This effect is limited, however, since a $v_{bo,9}\sim 3$ means a shocked proton thermal energy of $\frac{3}{16}m_pv^2=0.4v_9^2m_ec^2\sim m_ec^2$, such that the number of pairs produced from the shock thremal energy is at most comparable to the number of shocked electrons.

\section{Summary}\label{sec:conc}

We derive the spectrum and luminosity observed from a fast, $> 30,000$ km/s SN collisionless shock propagating through an optically thick wind, and find that the injection of soft photons into the interaction region plays the main role in shaping the spectrum and light curve. While previous works focused at lower shock velocities, $\lesssim 10,000$ km/s, and provided only rough predictions regarding the observed spectrum \citep[e.g.][]{Chevalier_Irwin2012,Svirski+2012}, here we elaborate on these works and provide, for the first time, a detailed prediction for the observed spectrum. Fast SN shocks propagating through a thick WR wind develop a flat spectrum with an initial frequency range $0.1\lesssim T\lesssim 50$ keV that becomes wider with time. If a SN explosion of a red or a blue supergiant drives a fast shock and the shock traverses a thick wind, the spectrum depends on the ratio $\frac{R_*}{R_{bo}}$, which value cannot be constrained a-priori.

In detail, we solve for the observed signal as a function of $f_{inj}$, $T_{inj}$, $t_{bo}$ and $v_{bo}$. When the shock is fast cooling, $f_{inj}$ and $T_{inj}$ determine the spectrum at $\tau\gtrsim 1$, and assuming  a standard wind density profile ($\rho\propto r^{-2}$), $t_{bo}$ and $v_{bo}$ determine the luminosity evolution. First, we find the threshold conditions for a fast cooling of the shock, Equations \ref{eq:condition} and \ref{eq:condition2}. When these conditions are satisfied, solving Equations \ref{eq:energy_balance}, \ref{eq:slopes2} and \ref{eq:Tmin} provides the spectral slopes and ranges. Generally, at lower energies, the spectrum is a power-law $\nu F_{\nu}\propto \nu^{\alpha}$ of $0\lesssim\alpha<1$, and it typically has a break above which $\alpha=-f_{inj}$. The spectrum spans from $T_{min}\gtrsim T_{inj}$ up to $m_ec^2/\tau^2\lesssim T_{max}\lesssim T_h$, where the exact $T_{min}$, $T_{max}$ and the break temperature are found by solving the equations as described above. An injected photon number flux mildly above the fast cooling threshold yields a hard spectrum, while a higher flux yields a softer spectrum. Over a wide range of injection parameters, the spectrum that evolves is approximately flat, $\nu F_{\nu}=Const$. The condition for a flat spectrum is given by Equation \ref{eq:flat2}.
Section \ref{sec:picture} provides a brief description of the physical processes that take part in shaping the spectrum.
When propagating through a \emph{thin} wind, the interaction signal is sub-dominant but it provides a unique signature through its X-ray luminosity decay pattern, $L_X(t)\propto t^{-n}$, where $n$ is the number of collisions with the hot layer that brings an Optical photon up to the X-ray detection window.

\begin{figure*}
	\centering
        \epsscale{1} \plotone{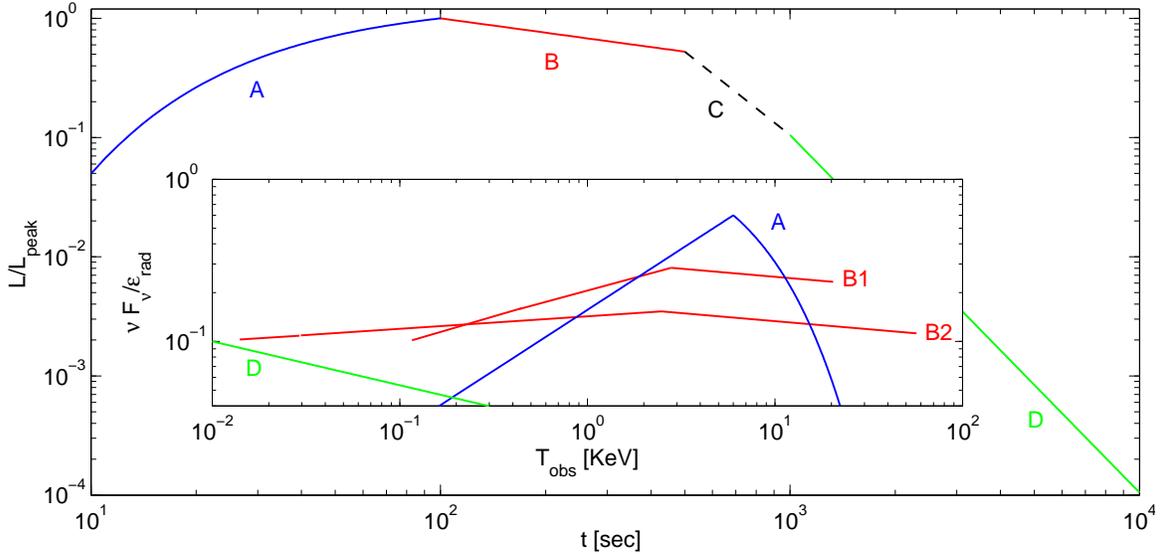}
	\caption{A schematic evolution of the X-ray luminosity (main figure) and the spectrum (inset) from a SN shock breakout through a thick WR wind, sketched for $t_{bo}=100$ s, $\tau_{bo}=5$, $f_{inj}=0.1$ and an initial $T_{inj}=0.1$ keV. (A) is the breakout pulse, with a non-thermal spectrum peaking at a few keV. (B) is the phase of IC fast cooling over injected photons, matching $\tau\gtrsim 1$, with $L\propto t^{-0.4}$ and a nearly flat spectrum that gradually widens from B1 to B2. (C) is the luminosity drop by $\sim 1/f_{inj}$ that follows the transition to a slow cooling at $\tau\lesssim 1$, and (D) reflects $\tau\ll 1$, with $L_X\propto t^{-3}$ and a negative spectral slope that steepens with time.}
	\label{fig:evolution}
\end{figure*}

After deriving our model we apply it to infer the observed spectrum and light curve from a SN shock breakout through a thick WR wind. Figure \ref{fig:evolution} depicts the different evolution stages of the X-ray luminosity and the spectrum for typical WR parameters, $t_{bo}=100$ s, $v_{bo,9}=6$, $\tau_{bo}=5$, $f_{inj}=0.1$ and an initial $T_{inj}=0.1$ keV. The first radiation observed is the breakout pulse, with a rise time $\sim 100$ s (A in Figure \ref{fig:evolution}) and a non-thermal spectrum peaking at a few keV (A in the inset). The collisionless shock that forms after the breakout is cooled efficiently and the bolometric luminosity decays slowly, $L\propto t^{-0.4}$ (B in Figure \ref{fig:evolution}). The observed spectrum is nearly flat, with an initial relatively narrow range in the X-rays (B1 in the inset) that gradually widens beyond X-rays as $\tau\rightarrow 1$ (B2 in the inset). At $\tau<1$ the shock cooling becomes slow, and the X-ray luminosity drops by a factor $\sim 1/f_{inj}$ within a dynamical timescale (C in Figure \ref{fig:evolution}). Later, when $\tau\ll 1$, the X-ray luminosity is dominated by $\lesssim$ keV photons and decays rapidly, $L_X\propto t^{-3}$ (D in Figure \ref{fig:evolution}).

The solution we provide for the thick wind phase is general and may apply to other physical settings involving an irradiated fast collisionless shock that propagates through an optically thick medium.

G.S. and E.N. were partially supported by an ISF grant (1277/13), an ERC starting
grant (GRB-SN 279369), and the I-CORE Program of the Planning and Budgeting Committee and The Israel Science Foundation (1829/12).

\appendix

\section{A fast IC cooling}\label{sec:fast cooling}
\subsection{A derivation of the velocity threshold}
Here we show that a shock propagating at $v>30,000$ km/s through a thick wind cannot cool efficiently without the aid of external radiation.
In \cite{Svirski+2012} we find the free-free cooling time of the hot layer at the breakout
\begin{equation}\label{eq:ff-cool}
t_{cool,bo}^{ff}\approx\frac{1}{25}t_{bo}v_{bo,9}^4T_{h,60}^{-1/2},
\end{equation}
which implies that for breakout shock velocities $v_{bo,9}> 2.25$ free-free emission fails to support a fast cooling already at the breakout. Replacing the radiation mediated shock compression factor of $7$ by a collisionless shock compression factor of $4$ implies that the gas behind the shock cannot cool fast by free-free emission if $v_{bo,9}> 2$. However, emission of soft photons by the unshocked upstream layer may still provide an efficient IC cooling. While the free-free emissivity of the upstream layer is lower than that of the shocked gas, due to its lower density, an upstream bound-free emissivity (irrelevant for the hot downstream) as high as ten times the free-free one would push the threshold velocity to $v_{bo,9}=2.5$. Since the threshold shock velocity has a rather small sensitivity to the actual cooling process and to the exact density, we adopt a $v_{bo,9}=3$ as a fiducial threshold velocity.

Unable to cool efficiently at $v_{bo,9}>3$, the cold layer electrons reach $T_h$, and emit $T_h$ photons that cannot cool the shock via IC. The soft tail of the free-free emission can neither provide the photon flux required for an efficient IC cooling: While a fast free-free cooling satisfies $f_{tail}\equiv \frac{L_{tail}}{L_{fast}}=\frac{T_{tail}}{T_h}$, a slow free-free cooling implies $f_{tail}<\frac{T_{tail}}{T_h}$, below the threshold defined in Equation \ref{eq:condition}.
Hence, in the absence of an external radiation, a $v_{bo,9}> 3$ implies a slow cooling and a single temperature $\sim T_h$ shared by the electrons of both the shocked layer and the upstream, and by the photons they emit.

\subsection{A slow IC cooling at $\tau<1$}
A fast shock cooling via IC,
over the radiation field that is produced by its own cooling,
can only last while $\tau>1$. The IC fast cooling condition, $\dot{\epsilon}_{IC}>\dot{\epsilon}_{gas}$, breaks when
\begin{equation}\label{eq:IC_cond}
\dot{\epsilon}_{IC}\approx\sigma_Tnc\epsilon_{gas}\frac{\tau v}{c}\frac{4T}{m_ec^2}
\approx\dot{\epsilon}_{gas}\approx\frac{\epsilon_{gas}}{t_{cool}}\approx\frac{\epsilon_{gas}}{t}
\end{equation}
where $t_{cool}=t$ corresponds to the transition from fast to slow cooling and $n$ is the electron number density. Substituting $\sigma_Tnvt=\sigma_TnR=\tau$, yields $\tau\approx\sqrt{\frac{m_ec^2}{4T}}\sim1$. In addition, at $\tau<1$ most photons that leave the hot layer never come back, implying a transition to slow IC cooling at $\tau\sim 1$.

\section{The cold shocked layer's compression}\label{sec:compression}

For typical WR parameters, the balance temperature of the cold layer electrons, $T_c$, is of order $1$ keV. Since protons temperature immediately behind the collisionless shock is $\frac{3}{16}m_pv^2\sim 200v_9^2$ keV, and since they lose their energy through Coulomb collisions within a fraction $\sim \left(\frac{v}{c}\right)^2$ of a dynamical time \citep{Katz+2011}, the downstream temperature quickly drops by a factor $\sim 100v_9^2$, i.e. by $3-4$ orders of magnitude for the relevant shock velocities. If the post-shock pressure is dominated by the gas, this temperature drop implies a compression of the cooled gas by the same factor.

At the density and temperature of the hot shocked layer, the breakout free-free to IC emissivity ratio is $\frac{\varepsilon^{ff,bo}}{\varepsilon^{IC,bo}}\propto n_eT_e^{-1/2}\approx4\cdot 10^{-2}v_{bo,9}^{-2}$ \citep{Svirski+2012}. A compression by $100v_9^2$ and a temperature drop from $T_h\sim 100$ keV to $1$ keV bring the emissivity ratio to $\approx40$. If free-free becomes the dominant electrons cooling process, a free-free cooling runaway may develop: A more efficient free-free cooling brings the electrons to a temperature $T<1$ keV, which implies, under a constant post-shock gas pressure, a further compression. This compression further enhances the free-free emissivity, and the gas continues to cool and compress until thermal equilibrium temperature, $\lesssim 50$ eV, is reached (implying an overall compression factor $\gtrsim 2000v_9^2$).

A cooling runaway of the cold shocked layer may lead to absorption of soft photons, which alters the spectrum that develops. However, a scenario involving a collisionless shock that propagates at $v_9\ge 3$ implies at least three different sources for non-thermal pressure, and it is very unlikely that neither of them gets to dominate the post-shock pressure and prevent a further compression and a runaway, when the gas is compressed by a factor $100v_9^2$. These sources are: Accelerated protons, the shock-induced magnetic field and the WR wind-induced magnetic field. We now discuss the contribution of each: (1) Accelerated protons: A modest energy fraction $\epsilon_p=(10v_9)^{-8/3}$ carried by accelerated protons, will dominate the post-shock pressure when compressed by $100v_9^2$, since the protons' energy density scales as $V^{-4/3}$, where $V$ is the gas volume, and the extra $V^{-1/3}$ reflects a $PdV$ work. Actual $\epsilon_p$ values in WR shocks are probably much higher, $\epsilon_p\sim 0.1$ \citep[e.g.][]{Chevalier_Fransson2006,Tatischeff+2009}, implying that the cooled shocked gas is only compressed by a factor of a few. This component probably dominates the post-shock pressure. (2) The shock-induced magnetic field: A modest such field, carrying a fraction $\epsilon_B=(10v_9)^{-2}$ of the shock energy, is amplified to dominate the pressure under a $100v_9^2$ compression. For WR shock velocities this requires $\epsilon_B\sim 10^{-3}-10^{-4}$, which is likely satisfied in WR shocks \citep{Chevalier+2006,Chevalier_Fransson2006}. (3) The WR wind-induced magnetic field: A shock density jump by $4$ and a further compression by $100v_9^2$ yield a magnetic energy density $\sim 10^{6}v_9^4B_{10}^2\,\rm{erg\,cm^{-3}}$, where $B_{10}=B/10\rm{G}$. For a magnetic field $B\sim 10$ G, this is comparable to the energy density deposited by the shock at the breakout (for a WR breakout radius of $10^{12}$ cm), $\sim3\times 10^6v_9^3\,\rm{erg\,cm^{-3}}$ (Equation \ref{eq:L(t)}). Such a field is likely induced by a WR wind regardless of the shock-induced field \citep[e.g.][]{Eichler+1993,Kholtygin+2011}.

\bibliographystyle{apj}
\bibliography{wind_breakout}
\end{document}